\documentclass{article}

\usepackage{arxiv}
\usepackage[utf8]{inputenc} 
\usepackage[T1]{fontenc}    
\usepackage{hyperref}       
\usepackage{url}            
\usepackage{booktabs}       
\usepackage{amsfonts}       
\usepackage{nicefrac}       
\usepackage{microtype}      
\usepackage{lipsum}
\usepackage{graphicx}
\usepackage{tikz}
\usepackage{placeins}
\usepackage{multirow}
\usepackage{mathtools}
\usepackage{amsmath}
\usepackage{mathpazo} 
\usepackage{amssymb}
\usepackage{floatrow}
\usepackage{soul}
\usepackage{algpseudocode}
\usepackage{algorithmicx}
\usepackage{algorithm}
\usepackage{pgfplots}

\graphicspath{ {./images/} }

\title{The Future of Document Indexing: GPT and Donut Revolutionize Table of Content Processing}

\author{
    Degaga Wolde Feyisa \\
      Dreeven Technologies Inc, St-Lambert, QC, Canada \\
      \texttt{degagawolde@gmail.com} \\
       \and
     Haylemicheal Berihun \\
      Dreeven Technologies Inc, St-Lambert, QC, Canada \\
      \texttt{haylemicheal.mekonnen@gmail.com} \\
      \and
     Amanuel Zewdu \\
      Dreeven Technologies inc. \\
      \texttt{amanuelzewdu21@gmail.com} \\
       \and
     Mahsa Najimoghadam \\
      Dreeven Technologies Inc, St-Lambert, QC, Canada \\
      \texttt{naji.mahsa@gmail.com} \\
       \and
      Marzieh Zare\\
      Dreeven Technologies Inc, St-Lambert, QC, Canada \\
      \texttt{marzieh.zare@Dreeven.com} \\
  }
  
\begin{document}
\maketitle
\begin{abstract}
Industrial projects rely heavily on lengthy, complex specification documents, making tedious manual extraction of structured information a major bottleneck. This paper introduces an innovative approach to automate this process, leveraging the capabilities of two cutting-edge AI models: Donut, a model that extracts information directly from scanned documents without OCR, and OpenAI GPT-3.5 Turbo, a robust large language model. The proposed methodology is initiated by acquiring the table of contents (ToCs) from construction specification documents and subsequently structuring the ToCs text into JSON data. Remarkable accuracy is achieved, with Donut reaching 85\% and GPT-3.5 Turbo reaching 89\% in effectively organizing the ToCs. This landmark achievement represents a significant leap forward in document indexing, demonstrating the immense potential of AI to automate information extraction tasks across diverse document types, boosting efficiency and liberating critical resources in various industries.
\end{abstract}


\section{Introduction}\label{section: intro}

Traversing extensive and intricate documents and locating essential information buried within numerous pages demands considerable time and effort. This inefficiency not only impedes understanding but also results in significant costs associated with manual data entry and extraction. To address this challenge, researchers have delved into the expanding realm of AI, with a particular emphasis on automating the extraction of information from large documents. Documents, whether in traditional written or electronic form, are essential for conveying information in various domains such as technology and business. Electronic documents, including PDF files, Microsoft Word documents, spreadsheets, emails, invoices, and presentations, play diverse roles like record-keeping, communication, collaboration, and supporting legal and financial transactions. The PDF format, introduced by Adobe Inc. in the 1990s, is particularly noteworthy for its widespread use in online document distribution, preserving original layout and formatting across devices and software. PDF documents, generated from applications like Microsoft Word and Adobe Acrobat, can encompass text, images, tables, graphs, hyperlinks, annotations, and forms, making them integral to modern document sharing and viewing practices. Additionally, a Table of Contents (ToC) acts as a guide in documents, facilitating quick navigation to specific sections of interest without the need to scan the entire text. Typically positioned after cover pages, the ToC includes main headings, subheadings, or subsections, enhancing document accessibility for readers.

The benefits of automating information extraction are multifold. First, it provides a concise overview of the document's structure and essential content, acting as a roadmap for efficient navigation. This is particularly critical in technical documents like construction specifications, where specific sections hold information on distinct components like electrical work or fire protection systems \cite{hong2022bros}. Second, it significantly reduces the time and costs associated with manual data extraction. Imagine the human resource burden of manually extracting key information from a multi-page contract compared to an AI system performing the task swiftly and accurately. These advantages have spurred considerable research and development in the field, with numerous companies offering visual document understanding tools (e.g., DocuVision, Adobe Experience Manager) \cite{elwany2022deeperdive, googledai, hypersciencedai, uipathdai}, and academic papers exploring various AI-based information extraction approaches \cite{hong2022bros, gralinski2020kleister, stanislawek2021kleister, lin2021vibertgrid}.

Our work falls within this domain, focusing on parsing large PDF documents and structuring their content for effortless navigation. Specifically, we target the extraction of heading numbers, titles, and their subheadings, aiming to build comprehensive ToCs. This structured information, often stored in formats like JSON files or SQL databases, serves as a vital key to indexing the document easily. By providing users access to this structured data through a user-friendly dashboard, we aim to empower them to visually comprehend the document's content and navigate its depths with unprecedented ease.

Navigating complex, multi-page documents can be a difficult task. Often, these documents are divided into distinct sections, each focused on specific topics or types of information. For instance, a construction specification document may contain sections dedicated to plumbing systems, floor openings, or concrete specifications. Efficiently traversing these sections relies heavily on structured information, such as well-organized ToCs. This structured information serves as a roadmap, guiding readers to relevant passages and facilitating comprehension.

Recognizing the importance of structure in document navigation, our work focuses on automatically extracting heading numbers, titles, and their corresponding subheadings from lengthy PDF documents. This extracted information forms the foundation of a comprehensive ToC, effectively transforming the document into a structured data format. We store this structured data in readily accessible formats like JSON files or SQL databases, enabling further analysis and visualization. We developed a user-friendly dashboard to empower users with effortless exploration of this extracted information. Through this interface, users can readily visualize the document's content, delve deeper into specific sections, and gain a holistic understanding of the document's structure and main themes.

The remaining sections of the paper are organized as follows. In Section \ref{section: rw}, we explore related works, discussing their contributions and shortcomings. The methodology employed in this study is outlined in Section \ref{section: method}. Our findings and the discussion of results are presented in Section \ref{section: disc}. The paper concludes with Section \ref{section: con}, providing a conclusion, and Section \ref{section: fw}, which outlines potential avenues for future work.

\section{Related Work} \label{section: rw}

Understanding document images is a critical task but presents significant challenges due to the need for complex functions such as text reading and overall document comprehension. Current methods for Visual Document Understanding (VDU) rely on off-the-shelf Optical Character Recognition (OCR) engines \cite{sayallar2023ocr, burchardt2023searches, akanksh2023automated, li2022pp, douzon2022improving, krieger4386107automated} for text extraction and focus on understanding tasks using OCR outputs. However, OCR-based approaches have several drawbacks, including computational costs, inflexibility in handling different languages or document types, and error propagation issues. To address these challenges, a few OCR-free VDU models have been proposed \cite{dhouib2023docparser, kim2022ocr}. In contrast to OCR-based models, OCR-free VDU models are designed to understand the visual content of documents without relying on traditional OCR techniques. These models use deep learning architectures, such as convolutional neural networks (CNNs) and transformers, to extract features from the visual content of documents. Large language models like GPT \cite{achiam2023gpt} can extract structured information from raw text. In the following sections, let's explore in-depth the literature on OCR-based and OCR-free models proposed by various authors for document understanding.

\subsection{OCR-Based VDU}

Li, et al. \cite{li2022pp} proposed an upgraded version of a document analysis system called PP-StructureV2. This new version contains two subsystems: Layout Information Extraction and Key Information Extraction (KIE). The focus of KIE is to extract specific information that users are interested in, and it includes subtasks such as Semantic Entity Recognition (SER) and Relation Extraction (RE). The paper also introduces several models that integrate text and layout information to improve the KIE process, including LayoutLM, LayoutLMv2, LayoutXLM, and XY-LayoutLM. Overall, the contribution of this paper is an improved document analysis system with enhanced functionality for extracting key information from unstructured documents.

PP-StructureV2 is one of the core components of PaddleOCR. Specifically, PP-StructureV2 provides layout analysis and key information extraction capabilities to PaddleOCR. The PaddleOCR team has developed an ultra-lightweight OCR system called PP-OCR, which prioritizes accuracy and speed for OCR industry applications \cite{du2020pp}. PP-OCRv3 \cite{li2022ppocr3} is an upgraded version of PP-OCRv2, which incorporates nine optimization strategies for text detection and recognition models. In comparison to PP-OCRv2, PP-OCRv3 demonstrates an 11\% improvement in English model precision, a 5\% improvement in Chinese model precision, and an average recognition accuracy improvement of over 5\% for 80 multilingual models, while maintaining a similar speed.

In paper\cite{douzon2022improving, krieger4386107automated}, the authors proposed the use of LayoutLM \cite{xu2020layoutlm}, a language model pre-trained on business documents, along with two new pre-training tasks and a post-processing algorithm to improve information extraction performance on expense receipts, invoices, and purchase orders. The proposed method significantly improves extraction performance on both public and private datasets. The authors used OCR (Optical Character Recognition) to extract textual and positional information from the business documents in the Business Documents Collection. This extracted information was then used as input for their pre-training language models on business documents. The accuracy of OCR is important because it affects the quality of data that goes into training these models, which can ultimately impact their performance in extracting relevant information from similar types of documents

\subsection{OCR-Free VDU}

Current document understanding models, such as LayoutLM, will often require OCR processing to extract the text from documents before they can be processed. While OCR can be an effective way to extract text from documents, it is not without its challenges. OCR accuracy can be impacted by factors such as the quality of the original document, the font used, and the clarity of the text. Furthermore, OCR is slow and computationally intensive which adds another layer of complexity. To overcome these challenges, new approaches are needed that can accurately interpret documents without the need for OCR. 

One popular approach to OCR-free VDU is to use end-to-end models that incorporate large language models, such as BERT (Bidirectional Encoder Representations from Transformers) \cite{dhouib2023docparser}, GPT(Generative Pre-trained Transformer), and their variants \cite{ul2012ocr}. These language models are pre-trained on large amounts of text data, enabling them to learn representations that capture the semantic relationships between words and phrases. By incorporating these language models into OCR-free VDU models, they can better understand the context and meaning of the visual content of documents, such as invoices, receipts, and forms. This approach has shown promising results in tasks such as information extraction, document classification, and question answering.

The Donut model \cite{kim2022ocr} is an end-to-end VDU solution that uses an encoder-decoder transformers model architecture, proposed by researchers from Naver CLOVA and recently made available to use with the HuggingFace transformers library. It encodes an image (split into patches using a Swim Transformer) into token vectors it can then decode, or translate, into an output sequence in the form of a data structure (which can then be further parsed into JSON) using the BART decoder model, publicly pre-trained on multilingual datasets.

\subsection{Large Language Models}

Large language models (LLMs), including transformer-based models like BERT \cite{devlin2018bert} or Megatron \cite{shoeybi2019megatron}, represent a subset of AI specifically designed for extracting crucial information from documents. Trained on extensive datasets encompassing both text and code, these models excel in identifying and extracting various entities from text, such as names, addresses, dates, and numbers. This adaptability positions them as valuable tools for extracting key information from diverse documents, such as invoices, contracts, and medical records \cite{mandvikar2023augmenting}.

Leveraging their capacity to grasp contextual intricacies within text, Language Models (LLMs) demonstrate superior accuracy compared to conventional methods. The proficiency of LLMs in comprehending context enhances the precision of entity identification. Furthermore, these models can be efficiently scaled to handle substantial volumes of documents, rendering them highly suitable for various applications. Despite their advantages, it is crucial to acknowledge the potential challenges associated with using LLMs for information extraction. Biases, factual errors, and limitations in understanding complex information can pose significant hurdles \cite{huang2023surveyhl, hadi2023survey}. Acknowledging and addressing these challenges is essential for ensuring the responsible and accurate deployment of LLMs in information extraction tasks. Recent advancements in LLM-based information extraction further enhance their utility. Techniques like multi-task learning and integration with additional resources, such as knowledge graphs, contribute to improved performance and expanded capabilities \cite{asai2023self,yang2023chatgpt,jeong2023study,krishna2023prompt}. These developments exemplify the dynamic nature of LLM research and its continuous evolution.

For further insights into challenges and methodologies related to LLM-based document information extraction, the work of Yousefi Maragheh et al., \cite{maragheh2023llm} can provide valuable perspectives. Additionally, \cite{xu2023large} provides a comprehensive survey of LLM applications in information extraction.

\section{Methodology} \label{section: method}

\subsection{Data Preparation}\label{dprep}
Any machine learning project requires a huge amount of data. Fortunately, the model used for this project, the Donut model, is pre-trained with a big dataset for both the image classification and information extraction tasks. So we fine-tuned it with our annotated image datasets to make it suitable for the two tasks. 

\begin{figure}[!ht]
    \centering
    \includegraphics[width=12cm]{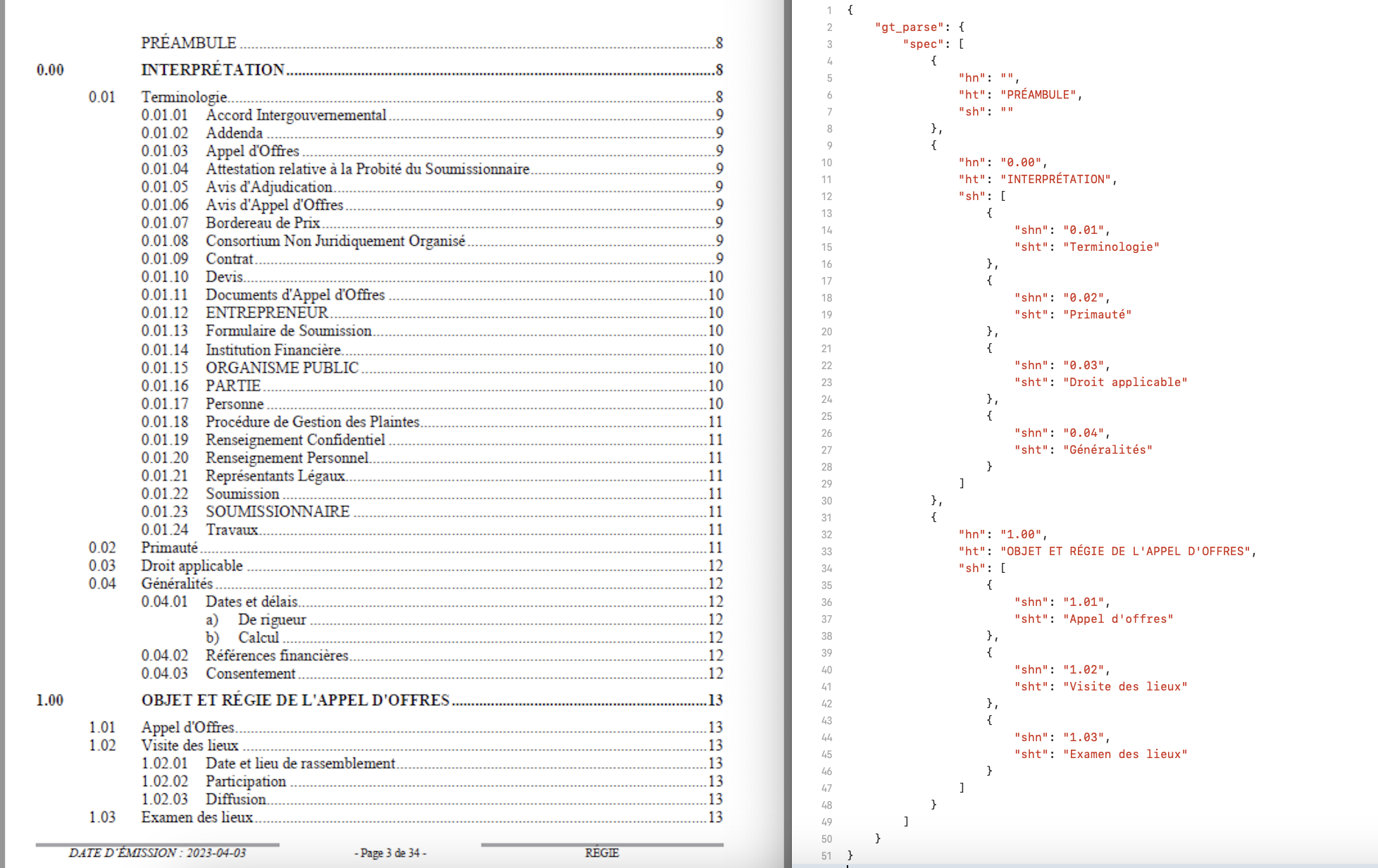}
    \caption{For this type of document, the first heading (h) is referred to as the heading, while the second heading (sh) is referred to as the subheading. The heading number (hn) comes before the heading title (ht). The same is true for the subheading number (shn) and title (sht).}
    \label{fig:oldformat}
\end{figure}

\begin{figure}[h]
    \centering
    \includegraphics[width=12cm]{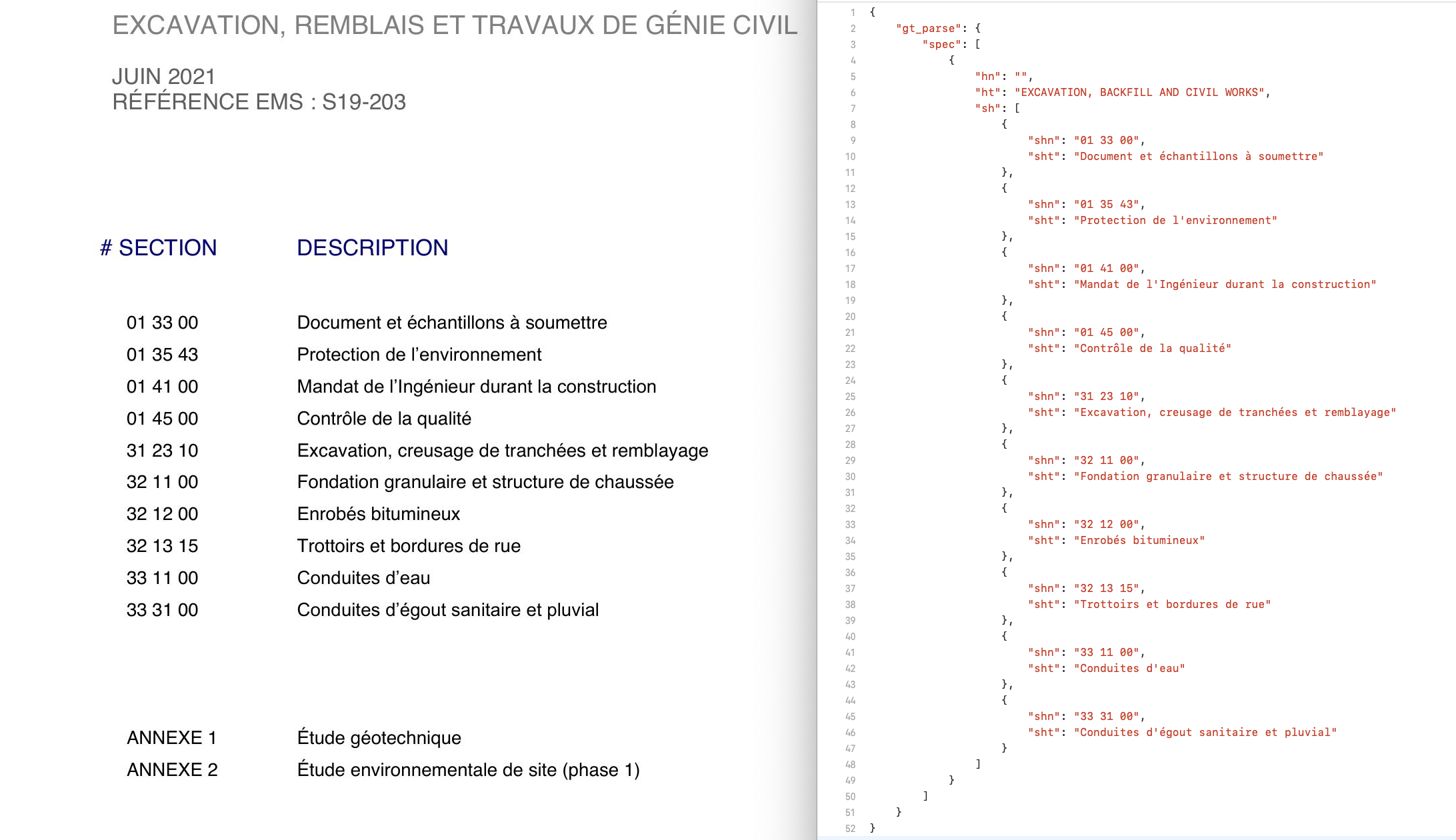}
    \caption{For this type of document, the division is referred to as the heading(h), while the sections are referred to as the subheading (sh). The heading number (hn) comes before the division title (ht). The same is true for the section number (she) and title (sht)}
    \label{fig:masterformat}
\end{figure}

The dataset used to fine-tune the image classifier model was composed of two classes of images taken from the specification document. The two classes are labeled as "ToC", and "other". To train the second model, which extracts heading and subheading information from the table of content pages identified by the first model, we used annotated "ToC" images extracted from varieties of specification documents. The data needed to fine-tune the Donut model should include the document's image and a JSON file containing the key information from the document. Figure \ref{fig:oldformat} and \ref{fig:masterformat} show how the data is prepared and annotation is generated for each ToC.

\FloatBarrier

Our dataset consisted of 200 annotated data samples utilized for both training and testing. We allocated 90\% of the dataset for training purposes and the remaining 10\% for testing. The images in the dataset were standardized to a height of 1260 and a width of 960 pixels. This ensured uniform dimensions for each image before being processed by the models in our study.

The dataset for the classification model is a page of the specification document with their respective class(table of contents, and other pages). In general, there are two classes: Table of Contents -- a table of contents for the document, mostly found in the first few pages, and other Pages -- pages that are not the table of contents. It can include white pages, cover pages, and other pages of the document that are not the main table of contents of the document.

\subsection{The Proposed AI-Based Document Indexing Technique}

\begin{figure}[!ht]
    \centering
    \includegraphics[width=14cm]{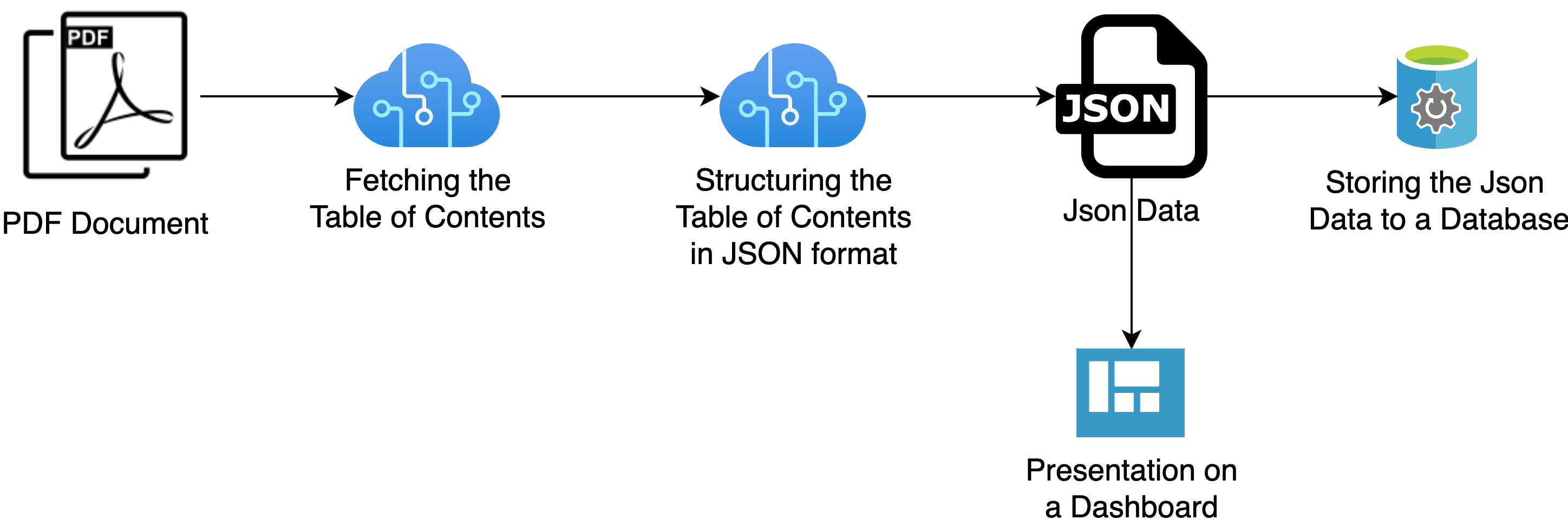}
    \caption{The proposed approach for retrieving and structuring table of contents of a PDF document.}
    \label{fig:method}
\end{figure}

In this section, we will discuss the methodology and process used for structuring a ToCs in a large PDF document. Figure \ref{fig:method} illustrates the proposed pipeline, which we will describe in detail. To begin with, the PDF document is uploaded, and each page is converted into an image. Subsequently, a classifier model is employed to categorize the images into one of two groups: tables of contents, or other pages. The pages classified as tables of contents are then fed into another model, which extracts information in  ToCs such as heading numbers, heading titles, subheading numbers, and subheading titles from each page. The extracted information is then structured and saved in a JSON file, which is parsed and displayed in the frontend application for the user to view. Additionally, the JSON file is stored in a database for future reference.

\subsubsection{Fetching the Table of Contents}
Document classification is a process of assigning categories or classes to documents to make them easier to manage, search, filter, or analyze. A document in this case is an item of information that has content related to some specific category. In a document classification application, an incoming stream or a set of documents is compared to a predefined set of rules. When a document matches one or more rules, the application performs some action. In this work classify a page of specification documents into a table of contents and other pages. Since we extract section titles and numbers from the table of contents of the documents, this is a very vital step toward the end objective.

\begin{figure}[!ht]
    \centering
    \includegraphics[width=14cm]{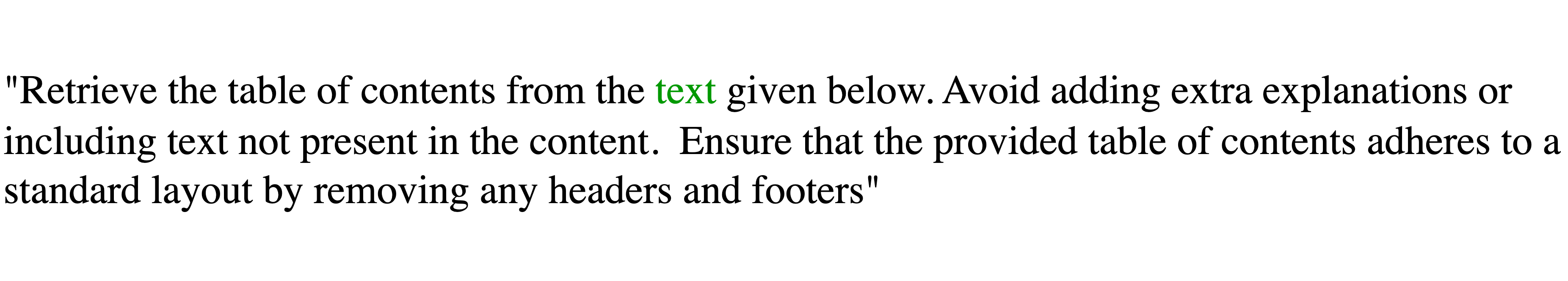}
    \caption{A prompt to extract ToC text from the raw text of a PDF file.}
    \label{fig:tocprompt}
\end{figure}

In the case of the Donut model, the process involves converting the pages into images, which are then input into the model to identify the ToC. Donut base model pre-trained on RVL-CDIP \cite{harley2015evaluation, larson2023evaluation} is pulled from hugging face and fine-tuned with classification data prepared as described in Section \ref{dprep}. On the other hand, when working with the OpenAI GPT-3.5 Turbo, a different approach is taken. Instead of converting the pages into images, a prompt provided in Figure \ref{fig:tocprompt} is utilized to retrieve the text of the ToC after obtaining raw text data from the PDF. This method underscores the versatility of the GPT-3.5 Turbo, showcasing its capability to handle textual information directly without the need for image conversion.

\subsubsection{Structuring the Table of Contents in JSON format}
Extracting data from unstructured documents is always a challenge. Previously we used to have rule-based approaches to tackle such problems. However, due to the nature of the rule-based mechanism, external knowledge sources, and manpower are required. To solve such issues, NLP is always a go-to solution for everyone. Deep learning has revolutionized the NLP field and to add to its hugging face has always delivered state-of-the-art solutions for multiple problems in NLP. We’re going to discuss one of the state-of-the-art OCR free visual document understanding called Donut and the OpenAI GPT-3.5 Turbo.

In this work, a Donut model pre-trained on the ICDAR-SROIE \cite{huang2019icdar2019} dataset is used. The model is first pulled from the hugging face repository and fine-tuned with our dataset. We sued 90\% for the training set and 10\% for the testing set. Two distinct types of specification documents exist in this context. The corresponding images were then annotated based on the desired JSON outputs. Following the data preparation outlined in \ref{dprep}, the data is supplied to the Donut model for the fine-tuning process.

\begin{figure}[!ht]
    \centering
    \includegraphics[width=14cm]{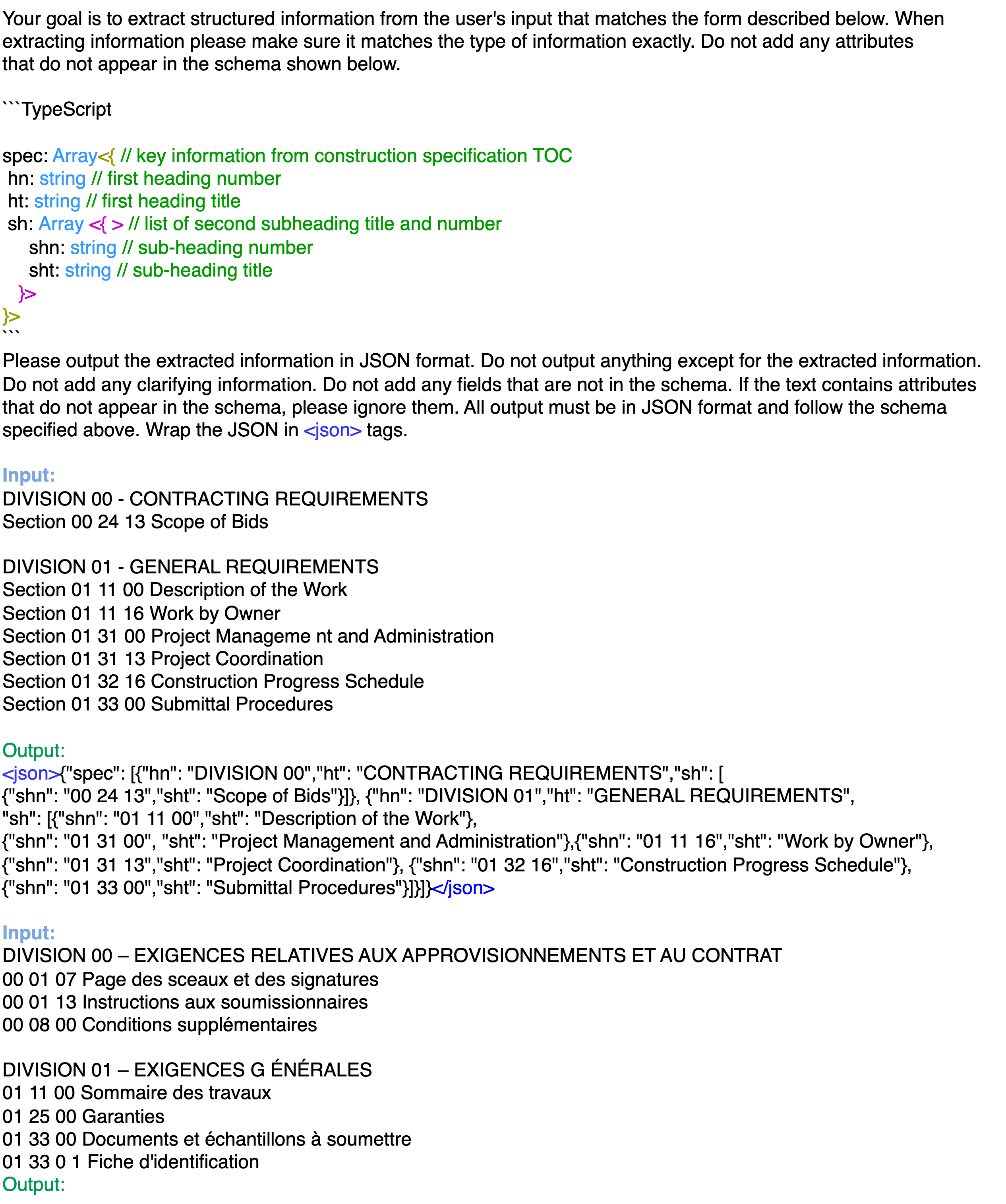}
    \caption{A prompt to extract key information based on a given example by formatting the output according to the provided schema.}
    \label{fig:prompt}
\end{figure}

For the OpenAI GPT-3.5 Turbo, we need to enhance the prompt design to leverage the capabilities of few-shot learning. This tex-to-text generation model takes a list of messages as input and generates a model-generated message as output. Few-shot learning involves providing the model with a small number of examples to enable it to understand and generalize from them. In our approach, we have incorporated a few-shot learning scenario by using a specific example in \autoref{fig:prompt}. This example serves as a guiding instance for the model to learn the desired behavior. By presenting the model with an illustrative example, we enable it to grasp the context and structure required for generating the desired output. This promotes a more efficient and effective learning process, allowing the model to adapt and generalize from the provided example to similar instances in the future.

The prompt is designed to instruct the model to extract JSON data based on the provided example and adhere to the given schema. This way, the few-shot learning paradigm enhances the model's ability to understand and respond to diverse prompts by learning from explicit examples, ultimately improving its performance in generating accurate and contextually relevant outputs.

\subsubsection{API integration}
After training our models and adding post-processing scripts, we exposed the pipeline using Flask API. These endpoints will be integrated with the front-end dashboard and yield the list of sections and divisions in a provided specification document.

\subsubsection{Presentation – Dashboard}

We developed a dashboard to showcase our work using Next.js, a React framework that provides additional features such as server-side rendering and the generation of static websites. React, a JavaScript library traditionally employed for building web applications rendered in the client's browser with JavaScript, is at the core of Next.js. Despite its popularity, this approach poses several challenges. These include issues with users who lack access to JavaScript or have disabled it, potential security concerns, significantly prolonged page loading times, and adverse effects on the site's overall search engine optimization. Developers have recognized these problems and seek solutions for a more inclusive and efficient web experience.

\subsection{Evaluation Metrics}
Assessing the effectiveness of a machine learning model requires the application of suitable evaluation metrics. In our scenario, we employ two distinct models: one designed for classification and another for organizing data within specified ToCs. The classification model's performance is gauged using accuracy. Meanwhile, for the second model, tasked with extracting structured data, we assess accuracy by comparing predicted key information with the corresponding ground truth JSON data. Exact matches are considered correct, while discrepancies are deemed missed key information. This approach allows us to compute the accuracy of structured data generation.

\subsection{Result and Discussion} \label{section: disc}
In this section, we discuss the results obtained after fine-tuning the model as described in the experimental setup. We evaluated the model using the above metrics and achieved an overall accuracy of 82.2\% by running it on 20 test documents and the results are shown in Figure 4. The accuracy of detecting heading numbers and titles is 90\% and 81\%, respectively. The accuracy for detecting subheading numbers and titles is 88\% and 79\%, respectively.


\begin{center}
	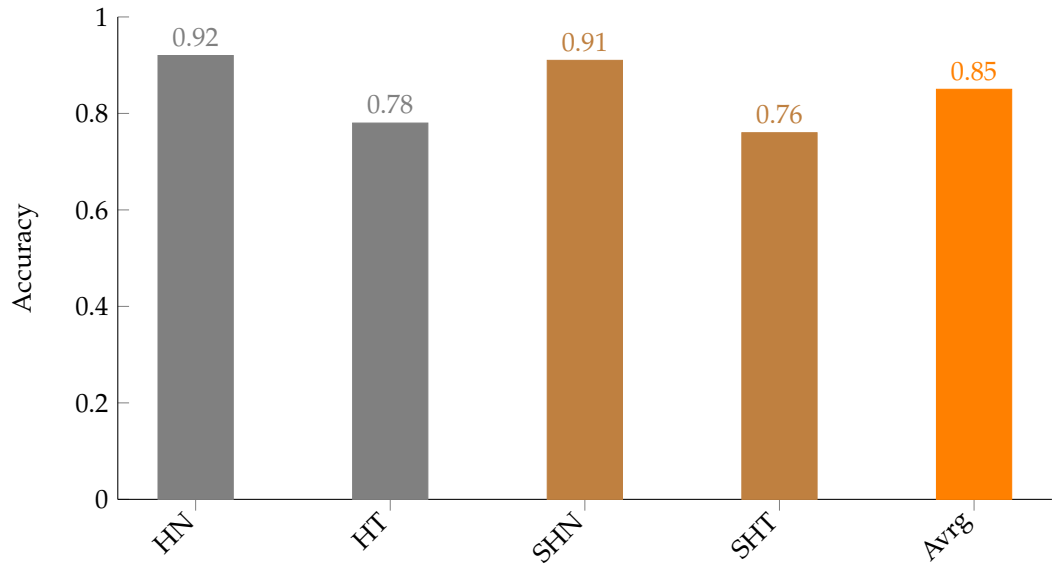
\begin{figure} [!ht]
		\begin{center}
			\begin{tikzpicture}
				\begin{axis}[
					ybar=-1cm,
					axis x line*=bottom,
					axis y line*=left,
					height=8cm, width=14cm,
					bar width=1cm,
					ymin=0, ymax=1,
					ylabel={Accuracy},
					symbolic x coords={HN, HT, SHN, SHT, Avrg },
					x tick label style={rotate=45, anchor=east, align=left},
					nodes near coords,
					nodes near coords align={vertical},
					xtick distance=1      
					]
					\addplot[gray,fill] coordinates {(HN,0.92)};	
					\addplot[gray,fill] coordinates {(HT,0.78)};
					\addplot[brown,fill] coordinates {(SHN,0.91)};
					\addplot[brown,fill] coordinates {(SHT,0.76)};				
					\addplot[orange,fill] coordinates {(Avrg,0.85)};
										          
				\end{axis}
			\end{tikzpicture}   
		\end{center}
			\caption{Peformance of Donut in detecting Heading number(HN), heading title(HT), subheading number(SHN), and subheading title(SHT).}
		\label{fig:searchegine2}
	\end{figure}
\end{center}

\begin{center}
	\begin{figure} [!ht]
		\begin{center}
			\begin{tikzpicture}
				\begin{axis}[
					ybar=-1cm,
					axis x line*=bottom,
					axis y line*=left,
					height=8cm, width=14cm,
					bar width=1cm,
					ymin=0, ymax=1,
					ylabel={Accuracy},
					symbolic x coords={HN, HT, SHN, SHT, Avrg },
					x tick label style={rotate=45, anchor=east, align=left},
					nodes near coords,
					nodes near coords align={vertical},
					xtick distance=1      
					]
					\addplot[gray,fill] coordinates {(HN,0.92)};	
					\addplot[gray,fill] coordinates {(HT,0.83)};
					\addplot[brown,fill] coordinates {(SHN,0.88)};
					\addplot[brown,fill] coordinates {(SHT,0.84)};				
					\addplot[orange,fill] coordinates {(Avrg,0.89)};
										          
				\end{axis}
			\end{tikzpicture}   
		\end{center}
			\caption{Peformance of OpenAI GPT-3.5 Turbo in detecting Heading number(HN), heading title(HT), subheading number(SHN), and subheading title(SHT).}
		\label{fig:searchegine1}
	\end{figure}
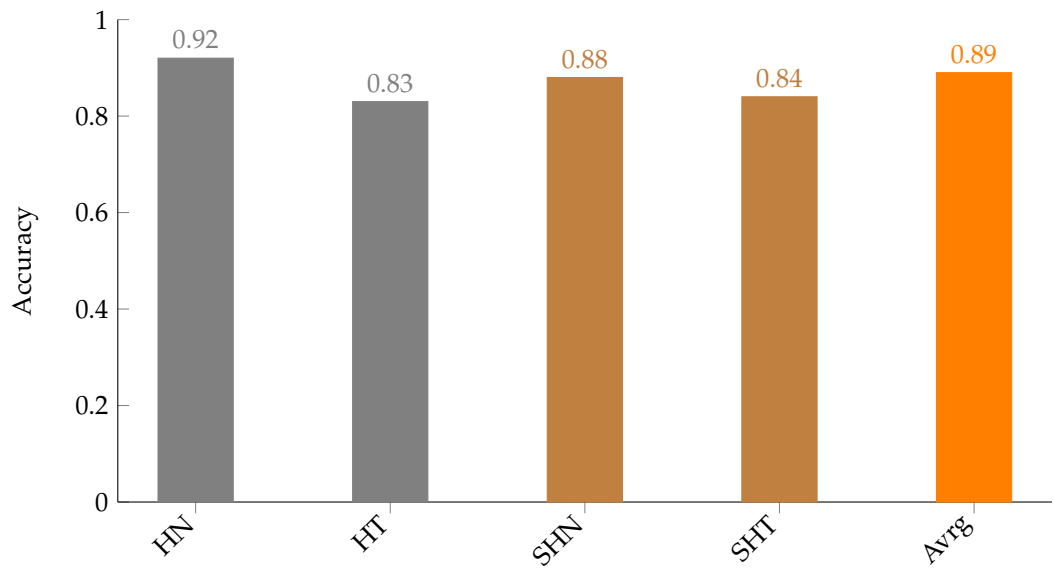
\end{center}
\FloatBarrier

\section{Conclusion} \label{section: con}
Failure to accurately structure information from large documents could create massive change orders, cost overruns, and schedule delays, which can negatively impact your bottom line. Large language models and computer vision are great tools to autonomously rearrange and categorize big documents like construction specification documents which may include various information about different technical sectors like architectural and mechanical work. By using an OCR-free document AI tool and OpenAI GPT-3.5 Turbo, we achieved an end-to-end tray that serves the list of sections and divisions in any specification document in a JSON.

\section{Future Work} \label{section: fw}
To further improve the accuracy of the document extraction model, several future tasks can be pursued. One such task is to collect more comprehensive data from a wider range of sources. This data could be used to refine the model's understanding of patterns. Another potential future task is to explore the use of more advanced machine learning algorithms, such as deep learning or reinforcement learning, to enhance the model's predictive capabilities. These algorithms could enable the model to identify more complex patterns and relationships in the data and to make more accurate predictions based on these insights. Finally, ongoing monitoring and evaluation of the model's performance will be critical to ensuring its continued accuracy and relevance over time. This could involve regular updates to the model's training data and ongoing testing and validation to identify any potential weaknesses or areas for improvement. By pursuing these future tasks, the ML model can continue to evolve and improve its predictions.

\bibliographystyle{unsrt}  
\bibliography{template}  

\end{document}